\documentclass{icrc2009}

\usepackage{graphicx}   
\usepackage{caption}    
\usepackage{fixltx2e}
\usepackage{multirow}
\usepackage{url}

\newcommand{\shorttitle}[1]%
{\markboth{Proceedings of the 31\MakeLowercase{$^{st}$} ICRC, 
{\L}\'{o}d\'{z} 2009}{#1} }
\newcommand{\etal}{\MakeLowercase{\textit{et al. }}} 

\begin{document}

\title{Highlight Talk: Recent Results from VERITAS
       \\ \vspace{-0.2cm}}

\author{\IEEEauthorblockN{
R.~A.~Ong\IEEEauthorrefmark{1},
V.~A.~Acciari\IEEEauthorrefmark{2}\IEEEauthorrefmark{22},
T.~Arlen\IEEEauthorrefmark{1},
T.~Aune\IEEEauthorrefmark{3},
M.~Beilicke\IEEEauthorrefmark{4},
W.~Benbow\IEEEauthorrefmark{2},\\
D.~Boltuch\IEEEauthorrefmark{5}, 
S.~M.~Bradbury\IEEEauthorrefmark{6},
J.~H.~Buckley\IEEEauthorrefmark{4},
V.~Bugaev\IEEEauthorrefmark{4},
K.~Byrum\IEEEauthorrefmark{7},
A.~Cannon\IEEEauthorrefmark{8},\\
A.~Cesarini\IEEEauthorrefmark{9},
L.~Ciupik\IEEEauthorrefmark{10},
Y.~C.~Chow\IEEEauthorrefmark{1},
P.~Cogan\IEEEauthorrefmark{11},
W.~Cui\IEEEauthorrefmark{12},
C.~Duke\IEEEauthorrefmark{13},\\
S.~J.~Fegan\IEEEauthorrefmark{1},
J.~P.~Finley\IEEEauthorrefmark{12},
G.~Finnegan\IEEEauthorrefmark{14},
P.~Fortin\IEEEauthorrefmark{15},
L.~Fortson\IEEEauthorrefmark{10},
A.~Furniss\IEEEauthorrefmark{3},\\
N.~Galante\IEEEauthorrefmark{2},
D.~Gall\IEEEauthorrefmark{12},
G.~H.~Gillanders\IEEEauthorrefmark{9},
S.~Godambe\IEEEauthorrefmark{14},
J.~Grube\IEEEauthorrefmark{8},
R.~Guenette\IEEEauthorrefmark{11},\\
G.~Gyuk\IEEEauthorrefmark{10},
D.~Hanna\IEEEauthorrefmark{11},
J.~Holder\IEEEauthorrefmark{5},
D.~Horan\IEEEauthorrefmark{7},
C.~M.~Hui\IEEEauthorrefmark{14},
T.~B.~Humensky\IEEEauthorrefmark{16},\\
A.~Imran\IEEEauthorrefmark{17},
P.~Kaaret\IEEEauthorrefmark{18},
N.~Karlsson\IEEEauthorrefmark{10},
M.~Kertzman\IEEEauthorrefmark{19},
D.~Kieda\IEEEauthorrefmark{14},
A.~Konopelko\IEEEauthorrefmark{20},\\
H.~Krawczynski\IEEEauthorrefmark{4},
F.~Krennrich\IEEEauthorrefmark{17},
M.~J.~Lang\IEEEauthorrefmark{9},
G.~Maier\IEEEauthorrefmark{11},
S.~McArthur\IEEEauthorrefmark{4},
A.~McCann\IEEEauthorrefmark{11},\\
M.~McCutcheon\IEEEauthorrefmark{11},
J.~Millis\IEEEauthorrefmark{21},
P.~Moriarty\IEEEauthorrefmark{22},
A.~N.~Otte\IEEEauthorrefmark{3},
D.~Pandel\IEEEauthorrefmark{18}, 
J.~S.~Perkins\IEEEauthorrefmark{2},\\
A.~Pichel\IEEEauthorrefmark{23}, 
M.~Pohl\IEEEauthorrefmark{17},
J.~Quinn\IEEEauthorrefmark{8},
K.~Ragan\IEEEauthorrefmark{11},
L.~C.~Reyes\IEEEauthorrefmark{16},
P.~T.~Reynolds\IEEEauthorrefmark{24},\\
E.~Roache\IEEEauthorrefmark{2},
H.~J.~Rose\IEEEauthorrefmark{6}, 
M.~Schroedter\IEEEauthorrefmark{17},
G.~H.~Sembroski\IEEEauthorrefmark{12}, 
A.~W.~Smith\IEEEauthorrefmark{7},
D.~Steele\IEEEauthorrefmark{10},\\
S.~P.~Swordy\IEEEauthorrefmark{16}, 
M.~Theiling\IEEEauthorrefmark{2},
S.~Thibadeau\IEEEauthorrefmark{4}, 
J.~A.~Toner\IEEEauthorrefmark{9},
A.~Varlotta\IEEEauthorrefmark{12},
V.~V.~Vassiliev\IEEEauthorrefmark{1},\\
S.~Vincent\IEEEauthorrefmark{14},
R.~G.~Wagner\IEEEauthorrefmark{7}, 
S.~P.~Wakely\IEEEauthorrefmark{16}, 
J.~E.~Ward\IEEEauthorrefmark{8},
T.~C.~Weekes\IEEEauthorrefmark{2},
A.~Weinstein\IEEEauthorrefmark{1},\\
D.~A.~Williams\IEEEauthorrefmark{3}, 
S.~Wissel\IEEEauthorrefmark{16},
M.~Wood\IEEEauthorrefmark{1},
and B.~Zitzer\IEEEauthorrefmark{12}}
\\
\vspace{-0.1cm}
\IEEEauthorblockA{\IEEEauthorrefmark{1} Department of Physics and
Astronomy, University of California, Los Angeles, CA 90095, USA}
\IEEEauthorblockA{\IEEEauthorrefmark{2} Fred Lawrence Whipple 
Observatory, Harvard-Smithsonian Center for Astrophysics, Amado, AZ 85645, USA}
\IEEEauthorblockA{\IEEEauthorrefmark{3} Santa Cruz Institute for Particle 
Physics and Department of Physics, University of California, \\
Santa Cruz, CA 95064, USA}
\IEEEauthorblockA{\IEEEauthorrefmark{4} Department of Physics, Washington 
University, St. Louis, MO 63130, USA}
\IEEEauthorblockA{\IEEEauthorrefmark{5} Department of  Physics and Astronomy 
and the Bartol Research Institute, University of Delaware, \\
Newark, DE 19716, USA}
\IEEEauthorblockA{\IEEEauthorrefmark{6} School of Physics and Astronomy, 
University of Leeds, Leeds, LS2 9JT, UK}
\IEEEauthorblockA{\IEEEauthorrefmark{7} Argonne National Laboratory, 
9700 S. Cass Avenue, Argonne, IL 60439, USA}
\IEEEauthorblockA{\IEEEauthorrefmark{8} School of Physics, University College 
Dublin, Belfield, Dublin 4, Ireland}
\IEEEauthorblockA{\IEEEauthorrefmark{9} School of Physics, National University 
of Ireland, Galway, Ireland}
\IEEEauthorblockA{\IEEEauthorrefmark{10} Astronomy Department, Adler 
Planetarium and Astronomy Museum, Chicago, IL 60605, USA}
\IEEEauthorblockA{\IEEEauthorrefmark{11} Physics Department, McGill 
University, Montreal, QC H3A 2T8, Canada}
\IEEEauthorblockA{\IEEEauthorrefmark{12} Department of Physics, Purdue 
University, West Lafayette, IN 47907, USA}
\IEEEauthorblockA{\IEEEauthorrefmark{13} Department of Physics, Grinnell 
College, Grinnell, IA 50112, USA}
\IEEEauthorblockA{\IEEEauthorrefmark{14} Department of Physics and Astronomy, 
University of Utah, Salt Lake City, UT 84112, USA}
\IEEEauthorblockA{\IEEEauthorrefmark{15} Department of Physics and Astronomy, 
Barnard College, Columbia University, NY 10027, USA}
\IEEEauthorblockA{\IEEEauthorrefmark{16} 
Kavli Institute for Cosmological Physics, University of Chicago, 
Chicago, Illinois 60637, USA}
\IEEEauthorblockA{\IEEEauthorrefmark{17} Department of Physics and Astronomy, 
Iowa State University, Ames, IA 50011, USA}
\IEEEauthorblockA{\IEEEauthorrefmark{18} Department of Physics and Astronomy, 
University of Iowa, Van Allen Hall, Iowa City, IA 52242, USA}
\IEEEauthorblockA{\IEEEauthorrefmark{19} Department of Physics and Astronomy, 
DePauw University, Greencastle, IN 46135, USA}
\IEEEauthorblockA{\IEEEauthorrefmark{20} Present address: Department of 
Physics, Pittsburg State University, 1701 South Broadway, \\ 
Pittsburg, KS 66762, USA}
\IEEEauthorblockA{\IEEEauthorrefmark{21} Present address:  Department 
of Physics, Anderson University, 1100 East 5th Street, Anderson, IN 46012}
\IEEEauthorblockA{\IEEEauthorrefmark{22} Department of Life and Physical 
Sciences, Galway-Mayo Institute of Technology, Dublin Road, Galway, Ireland}
\IEEEauthorblockA{\IEEEauthorrefmark{23} Instituto de Astronomia y Fisica del 
Espacio, Casilla de Correo 67 - Sucursal 28, (C1428ZAA) \\
Ciudad Autonoma de Buenos Aires, Argentina}
\IEEEauthorblockA{\IEEEauthorrefmark{24} Department of Applied Physics and 
Instrumentation, Cork Institute of Technology, Bishopstown, \\
Cork, Ireland \\
\vspace{-0.7cm}}
}

\shorttitle{R.A. Ong \etal Recent VERITAS Results} \maketitle

\begin{abstract}

VERITAS is a state-of-the-art ground-based gamma-ray observatory that operates in the
very high-energy (VHE) region  of 100 GeV to 50 TeV.
The observatory consists of an array of four 12m-diameter imaging 
atmospheric Cherenkov telescopes located in southern Arizona, USA.
The four-telescope array has been fully operational since 
September 2007, and over
the last two years, VERITAS has been operating with high efficiency 
and with excellent performance.
This talk summarizes the recent results from VERITAS, including the 
discovery of eight new VHE gamma-ray sources.

\end{abstract}

\begin{IEEEkeywords}
 gamma rays, galactic observations, extragalactic observations
\end{IEEEkeywords}

\section{Introduction}

Construction on the
VERITAS (Very Energetic Radiation Imaging Telescope Array
System) experiment, located at the basecamp of the
F.L. Whipple Observatory (FLWO) in southern Arizona, USA,
started in 2003 and was completed in
June 2007.  
As shown in Figure~1,
the array consists of four 12m-diameter imaging 
atmospheric Cherenkov telescopes, with
a typical baseline between telescopes
of $\sim$100\,m.
Each telescope has a
499-photomultiplier tube (PMT) camera, spanning
a field of view of $3.5^\circ$.
The signal from each camera pixel is amplified and recorded by a 
separate 500 MS/s Flash-ADC channel.
VERITAS employs a three-level trigger system;
Level 1 corresponds to constant fraction
discriminators on each pixel, 
Level 2 is a pattern trigger for each telescope,
and Level 3 is the array trigger.
More details on VERITAS can be found in 
\cite{VERITAS}.

Regular observations with the full four-telescope array
started in September 2007, with
approximately 1000 h per year of observations 
taken.
The array has operated extremely well during the
last two years; more than 95\% of the observations have at
least four telescopes operational.
The ability to take scientifically useful data under
partial moonlight was an important development -- it
adds approximately 30\% to the annual data yield.

With two years of data in hand, the performance attributes
of VERITAS are now well understood.
These attributes are an angular resolution
(68\% containment) of $< 0.1^\circ$,
a pointing accuracy of $< 50\,$arc-secs,
an energy range of 100\,GeV--50\,TeV, and
an energy resolution (above 200\,GeV) of 15--20\%.
The gamma-ray point source sensitivity of VERITAS
in its original configuration corresponds to 
the detection of a 1\% Crab Nebula source
at the five standard deviation ($\sigma$) level
in less than 50 h.
In Summer 2009, Telescope 1 was relocated to improve
the array geometry (see Figure~1).
This change had a significant impact on the performance
of VERITAS; a point source at the
1\% Crab Nebula flux level can now be detected in under
30 h of observation.

The observing time of VERITAS during its first two years
was divided into {\em Key Science Projects},
the {\em Bulk Program}, and
{\em Director's Discretionary Time}, corresponding to
50\%, 40\%, and 10\% of the observation time, respectively.
The Key Science Projects were a sky survey of the
Galactic plane in the Cygnus region,
a blazar program, a study of supernova remnants and pulsar
wind nebulae, and a search for dark matter.
The bulk program was determined by individual observing
proposals 
submitted to a time allocation committee (TAC).
The Director's Discretionary Time was set aside for GRB's,
ToO's, and occasional high-risk sources.

In this paper, we summarize the scientific highlights
of VERITAS, concentrating on recent results obtained in the
last year and presented at the 31st International
Cosmic Ray Conference (Lodz, Poland, July 2009).
For additional details, see the individual contributions
to the conference \cite{ICRC_VERITAS}.

 \begin{figure}[!t]
  \centering
  \includegraphics[width=3.0in]{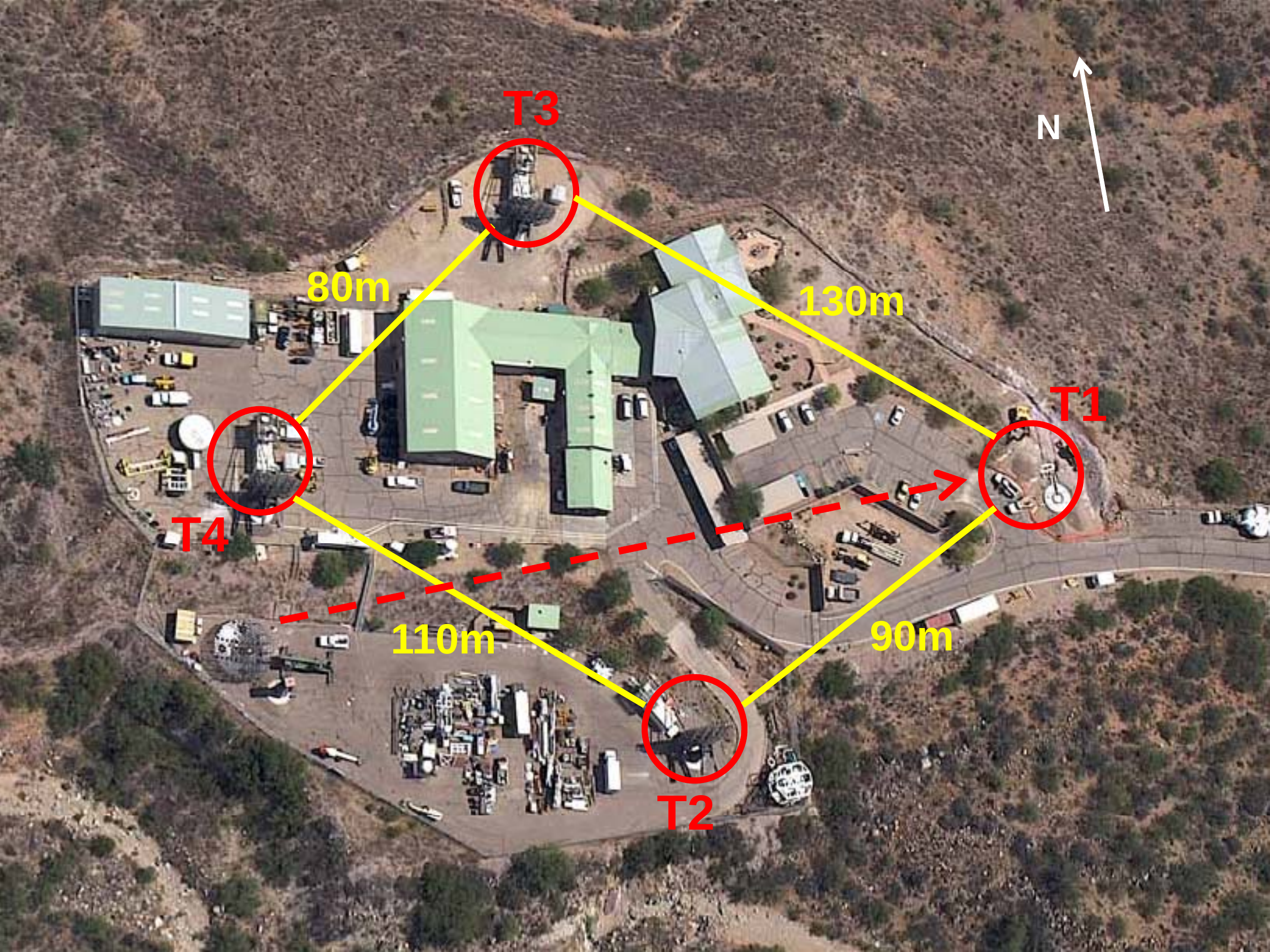}
  \caption{Aerial view of VERITAS at the Whipple Observatory basecamp
  in Arizona, USA.  The number of the telescopes specifies the order
  in which they were constructed; the yellow lines connecting
  telescopes indicate the inter-telescope baselines; the
  dashed red line indicates the relocation of Telescope 1
  that was carried out in Summer 2009.}
 \end{figure}

\section{Highlight: Detection of Starburst Galaxy M 82}

Starting with the discovery of Markarian 421 in 
1992 \cite{Punch},
atmospheric Cherenkov telescopes have been remarkably
successful in the detection of VHE gamma-radiation from
extragalactic sources.
Indeed, 
the VHE catalog presently includes
approximately 25 extragalactic entries \cite{TeVCAT}.
To date, however, all of these sources are
active galactic nuclei (AGN), mostly of the blazar type.
The observational data support the general picture in which
the VHE emission results from acceleration processes that
are ultimately powered by accretion onto a supermassive
black hole.
However,
the vast majority of galaxies are not active at the present
time, and an important question is whether there is
detectable VHE emission from galaxies not associated
with black hole activity.

The importance of finding new types of extragalactic
VHE emitters motivated the VERITAS observations of
the M 82 starburst galaxy.
M 82 is a prototype starburst galaxy where
the interaction of a group of galaxies produces
a very active starburst region.
In this region, a high cosmic ray density, as inferred
from radio synchrotron emission \cite{M82_radio}, 
is believed to result from
the high star formation and supernova (SNR) rate.
M 82 also contains a high mean gas density of $\sim$150
particles/cm$^2$ \cite{M82_gas}.
A natural mechanism to produce gamma rays involves the
interaction of cosmic rays (both hadrons and electrons)
with the dense gas and photon fields.
This is the very mechanism that produces the Galactic
diffuse gamma-ray emission in the Milky Way. 
Previous limits on the flux of VHE gamma rays from
M 82 $< 10$\% Crab Nebula flux have come
from Whipple \cite{M82_Whipple} and HEGRA \cite{M82_Hegra}.

The VERITAS M 82 data set, taken in dark time between 2007
and 2009, constitute a very deep exposure 
of 137 h.
Selection criteria to increase the sensitivity
of the instrument at high energies
(so called ``hard cuts'') were developed from
an {\em a priori} study of the Crab Nebula at
similar zenith angles as M 82.
These cuts yield a post-trials significance
of 4.8$\sigma$. 
The excess counts map
(see Figure~2) is consistent with a point source
at the position of M 82.
The detected gamma-ray flux of $\sim$0.9\% Crab Nebula
(E $>$ 700 GeV) is among the weakest VHE sources yet detected.
Numerous systematic checks were made 
to provide
confidence that the gamma-ray signal is genuine.
Complete details on the VERITAS detection of M 82 
can be found in a recent publication  
\cite{M82_VERITAS}.

The discovery of VHE gamma-ray emission from M 82 by
VERITAS represents the first detection of
gamma rays from a starburst galaxy
and the first extragalactic VHE source not clearly
associated with AGN activity.
The detected flux level is consistent
with theoretical predictions \cite{Persic,Torres}
that are based on standard mechanism of cosmic-ray
interactions.
At this meeting, we learned about the detection by HESS
of VHE emission from the  
starburst galaxy NGC 253 by HESS
\cite{NGC253_HESS} and, subsequently, about the 
detection by Fermi-LAT of high-energy gamma-ray
emission from M 82 \cite{M82_Fermi}.

\begin{figure}[!t]
  \centering
  \includegraphics[width=3.15in]{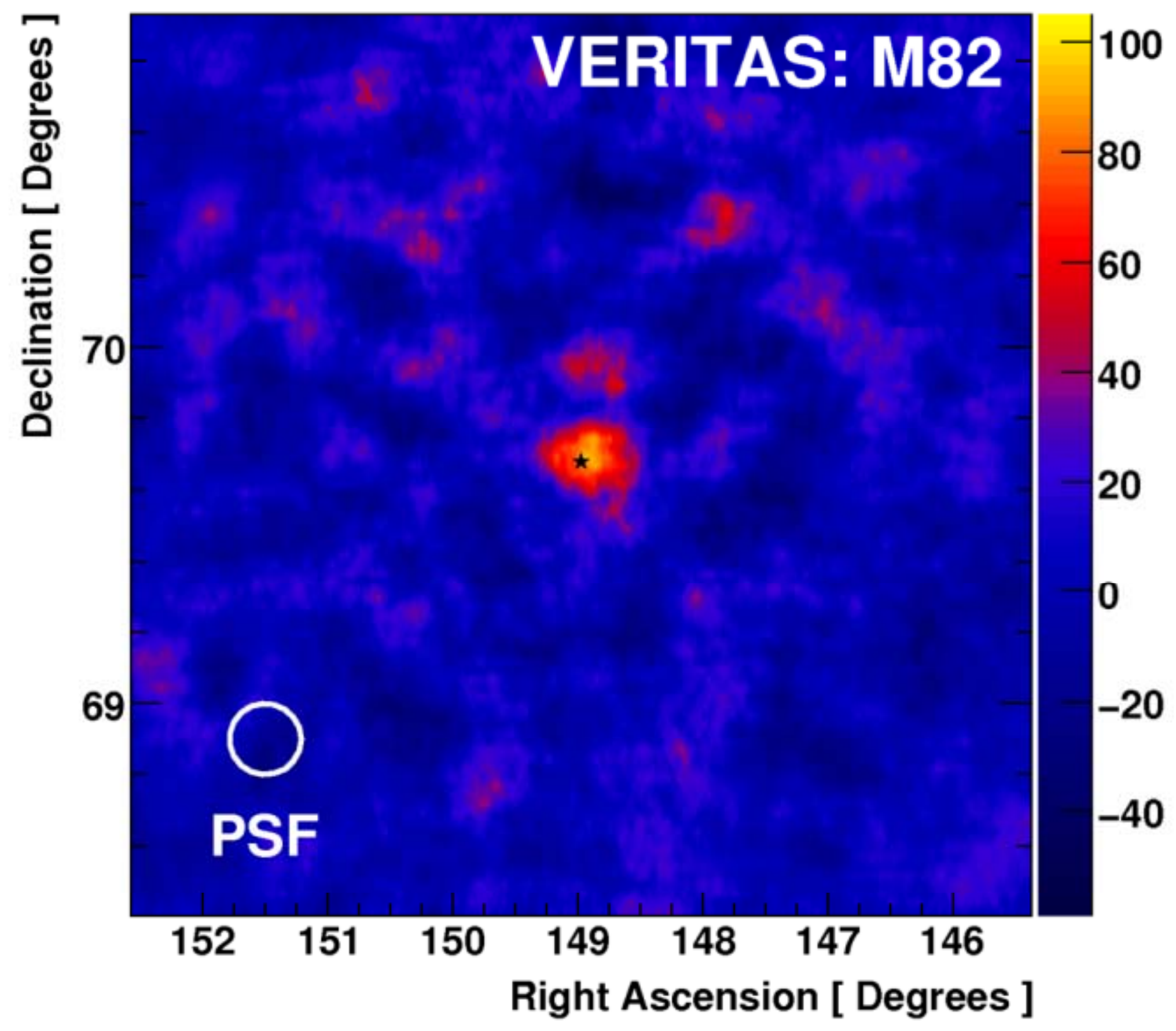}
  \caption{VHE gamma-ray image of M 82 by VERITAS.
           The image shows the measured excess
           of gamma-ray like events above the estimated
           background in a region centered on M 82. The
           white circle represents the point-spread
           function (68\% containment) of VERITAS and
           the black star indicates the location of the
           core of M 82.}
\end{figure}

\section{Highlight: Galactic Plane Survey}

The Sky Survey of the Galactic plane in the region of
Cygnus was a VERITAS key science project,
carried out over two observing seasons between 2007
and 2009 \cite{Weinstein}.
The Cygnus region is a natural target for a survey,
containing a variety of potential VHE sources, including
supernova remnants, pulsar wind nebulae, X-ray binaries,
and massive star clusters.
The first unidentified TeV gamma-ray source,
TeV J2032+4130,
was reported by HEGRA from a survey of Cygnus
\cite{Cygnus_HEGRA,TeV2032_HEGRA}.
At GeV energies,
Fermi-LAT has detected at least four distinct sources,
all associated with pulsars \cite{Cygnus_Fermi},
At $>$10 TeV,
Milagro reported two
unidentified sources
(MGRO J2031+41 and MGRO J2019+37) \cite{Cygnus_Milagro}.
MGRO J2031+41 appears to be associated with
TeV J2032+4130.

The VERITAS sky survey covers the region
of Galactic longitude $67^\circ < {\rm l} < 82^\circ$
and Galactic latitude $-1^\circ < {\rm b} < 4^\circ$
The observations consisted of a
base survey of 112 h and follow-up observations
of 32 h.
The base survey was carried out by a set of
grid pointings where grid points had
separations of $0.8^\circ$ in Galactic longitude
and $1.0^\circ$ in Galactic latitude.
Figure~3 shows an exposure-weighted map of the
base survey.
Observations were carried out using 
three and four VERITAS telescopes and at 
zenith angles less than $35^\circ$.
The exposure across the survey
region is relatively uniform with an effective
(acceptance-corrected) exposure of $\sim$6 h
for all points within the survey boundaries.

\begin{figure*}[!t]
  \centering
  \includegraphics[width=5.8in]{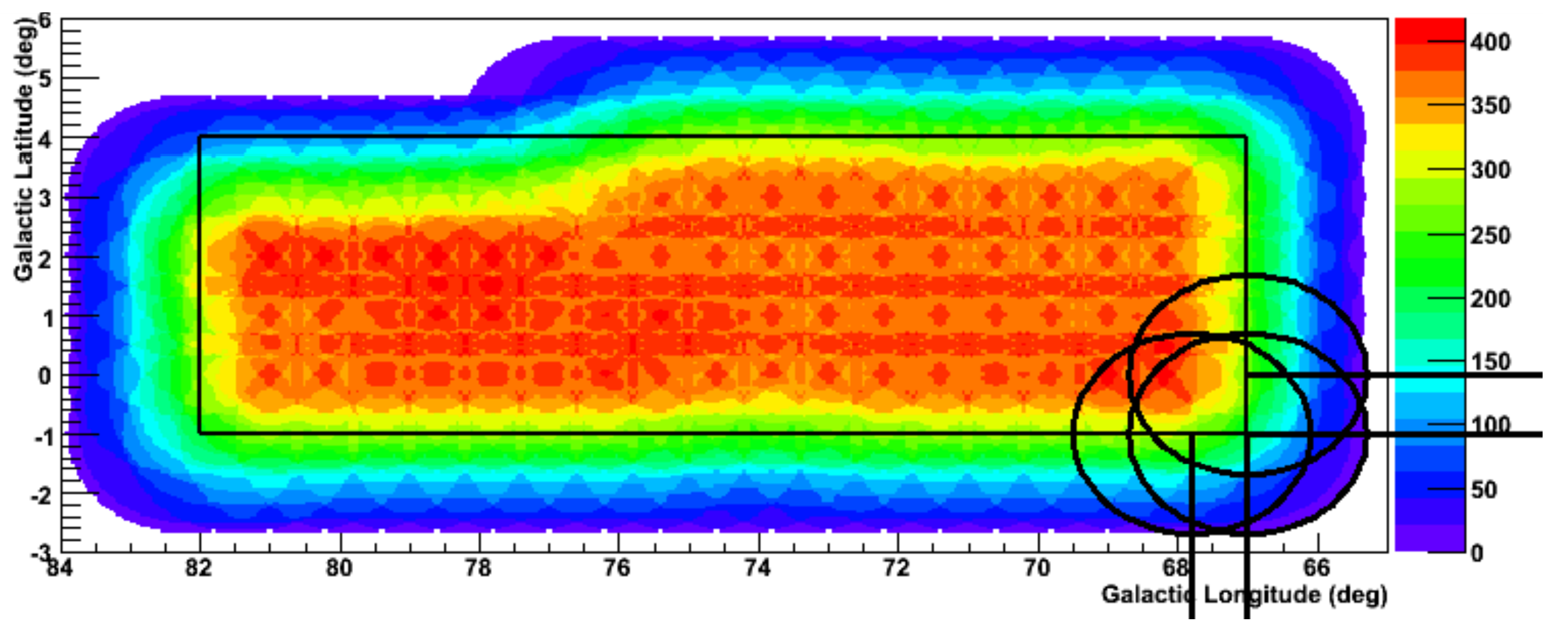}
  \caption{Exposure-weighted map of the VERITAS
  Sky Survey of the Cygnus region.  The color
  scale indicates the effective exposure in minutes.
  The black circles indicate the field-of-view
  of VERITAS at survey grid pointings separated
  by $1.0^\circ$ in Galactic longitude and
  $0.8^\circ$ in Galactic latitude.}
\end{figure*}

The analysis of the survey data was made using
a set of pre-defined selection criteria (cuts)
designed to maximize the sensitivity of the
survey to sources with varying 
spectral shape and extension.
In particular, cuts were made on the 
integrated charge (size) and on
the distance between the reconstructed direction
and direction of a potential source ($\theta^2$).
The cuts resulted in a total of four parallel
analyses ({\em i.e.} a trials factor of four);
see \cite{Weinstein} for details.

The preliminary result
from the base survey is that
no sources are detected at a significance
level greater than 5$\sigma$ (post-trials).
The survey sensitivity is estimated using a technique 
in which simulated gamma rays are injected
into background survey fields taken from actual data.
The estimated sensitivity leads to preliminary 99\% C.L. limits
on the flux of
VHE gamma rays of $< 3\%$ Crab Nebula (point source) 
and $< 8.5\%$ Crab Nebula 
(extended source of diameter $0.5^\circ$)
at a median energy of 200 GeV.
These limits are 3-4 times
more stringent
than those achieved in the previous work 
of HEGRA \cite{Cygnus_HEGRA}.
They also indicate that the population density of
VHE sources in the northern hemisphere
is markedly different than in the southern
hemisphere, where HESS found 12 sources above
a flux level of 5\% Crab Nebula \cite{Survey_HESS}.
Follow-up observations are continuing in 
specific regions of the survey.

\section{Extragalactic Sources}

\subsection{Blazars}

Until the detection of starburst galaxies,
the extragalactic sources detected at very
high energies were all active galactic nuclei (AGN),
of which blazars represented the dominant source
class.
The general picture for blazars involves
the accretion of matter onto a supermassive
black hole that powers relativistic jets of
plasma flow that are pointed in the direction
of Earth.
VHE particle acceleration takes place in the
jets, resulting in GeV and TeV gamma-ray emission.

The main science goal of studying blazars at
gamma-ray energies is to gain an understanding
of the physics taking place in jets and to
ultimately connect that physics to the
black hole accretion.
Another goal is to use the
gamma-ray emission from blazars as a probe
of intergalactic radiation fields, both
the extragalactic background light (EBL,
through the absorption process
$\gamma + \gamma \rightarrow e^+ e^-$)
and the intergalactic magnetic field.

Most high-energy blazars exhibit
``double-peaked'' spectral energy distributions
in which the low-energy peak can be attributed
to synchrotron emission and the high-energy peak
to inverse-Compton scattering.
However, 
as yet, we cannot conclusively pinpoint whether
the parent particles accelerated in the jet
are electrons or protons.
In this context, multiwavelength observations
can be particularly effective in constraining
model parameters.
To date almost all of the blazars detected
at TeV energies are high-frequency
peaked BL Lacertae (HBL) objects in which 
the synchrotron peak lies in the X-ray band.
This is to contrasted with low-frequency
peaked BL Lac (LBL) objects, predominantly detected
at GeV energies,  where the synchrotron
peak is in the radio band.

The VERITAS blazar key science project 
uses a multi-faceted approach to
improve our knowledge of the acceleration
and emission mechanisms taking place
in blazars.
The project is divided approximately equally
into discovering new sources, multiwavelength
campaigns, and targets of 
opportunity \cite{Benbow}.
Fifty blazars have been observed so far,
resulting in eleven detections and five discoveries
of VHE emission.
The first source discovered by VERITAS
was 1ES 0806+524, where a 40 hour exposure
in the 2007/2008 observing season 
led to a detection of this relatively
weak HBL at a flux level of
$\sim$2\% Crab Nebula \cite{1ES0806_VERITAS}.

A recent blazar discovery by VERITAS is
RGB 0710+591, an HBL at a redshift
of $z = 0.125$ \cite{RGB0710_VERITAS,Perkins}.
This source was detected by VERITAS at 
significance of $\sim 6 \sigma$
from 20 h of observations in 2009.
The relatively hard energy spectrum
(preliminary differential spectral index
$\Gamma = 2.8 \pm 0.3_{\rm stat}
\pm 0.3_{\rm sys}$) should provide
significant constraints on the density
of the EBL.

An important result from the VERITAS blazar program has
been the establishment of intermediate-frequency
peaked BL Lacertae (IBL) objects as emitters
in the VHE band.
IBL objects are thought of as an intermediate class
between LBLs and HBL's, although in reality it is likely
there is a continuum of objects.
The first IBL to be established at very high energies
was W Comae, a known EGRET source at
a redshift of $z = 0.102$, detected by VERITAS
during 40 h of observation in
Spring 2008 \cite{WComae_VERITAS}.
This source exhibited strong variability and a very
steep energy spectrum, with differential spectral
index of $\Gamma = 3.81 \pm 0.35_{\rm stat}
\rm 0.34_{\rm sys}$.
A second VHE flare from W Comae was detected in
June 2008 at a significantly higher flux level
than the first \cite{Maier}.

The second IBL to be discovered by VERITAS
was 3C 66A \cite{3C66A_VERITAS}.
This is a rather famous source 
that has long been considered a likely candidate
for VHE emission.
33 h of observation in 2008 by VERITAS
resulted in a strong detection
($\sim 21 \sigma$) of 3C 66A with variability
seen on day time scales. The measured
energy spectrum is very steep,
$\Gamma = 4.11 \pm 0.4_{\rm stat}
\pm 0.6_{\rm sys}$, which may be entirely
due to the absorption by the EBL;
3C 66A has an uncertain redshift of
$z = 0.44$.
3C 66A lies $0.12^\circ$ away from the radio
galaxy 3C 66B. MAGIC reported the detection
of VHE emission from the region that
is consistent
(at 85\% C.L.) with 3C 66B. However, as
shown in Figure~4, the VERITAS data exclude
the position of 3C 66B at the 4.3$\sigma$ level.
Fermi LAT has detected bright emission from
3C 66A; the results from a joint
Fermi-VERITAS study of the source are 
discussed elsewhere at this conference
\cite{Reyes}.

\begin{figure}[!t]
  \centering
  \includegraphics[width=3.15in]{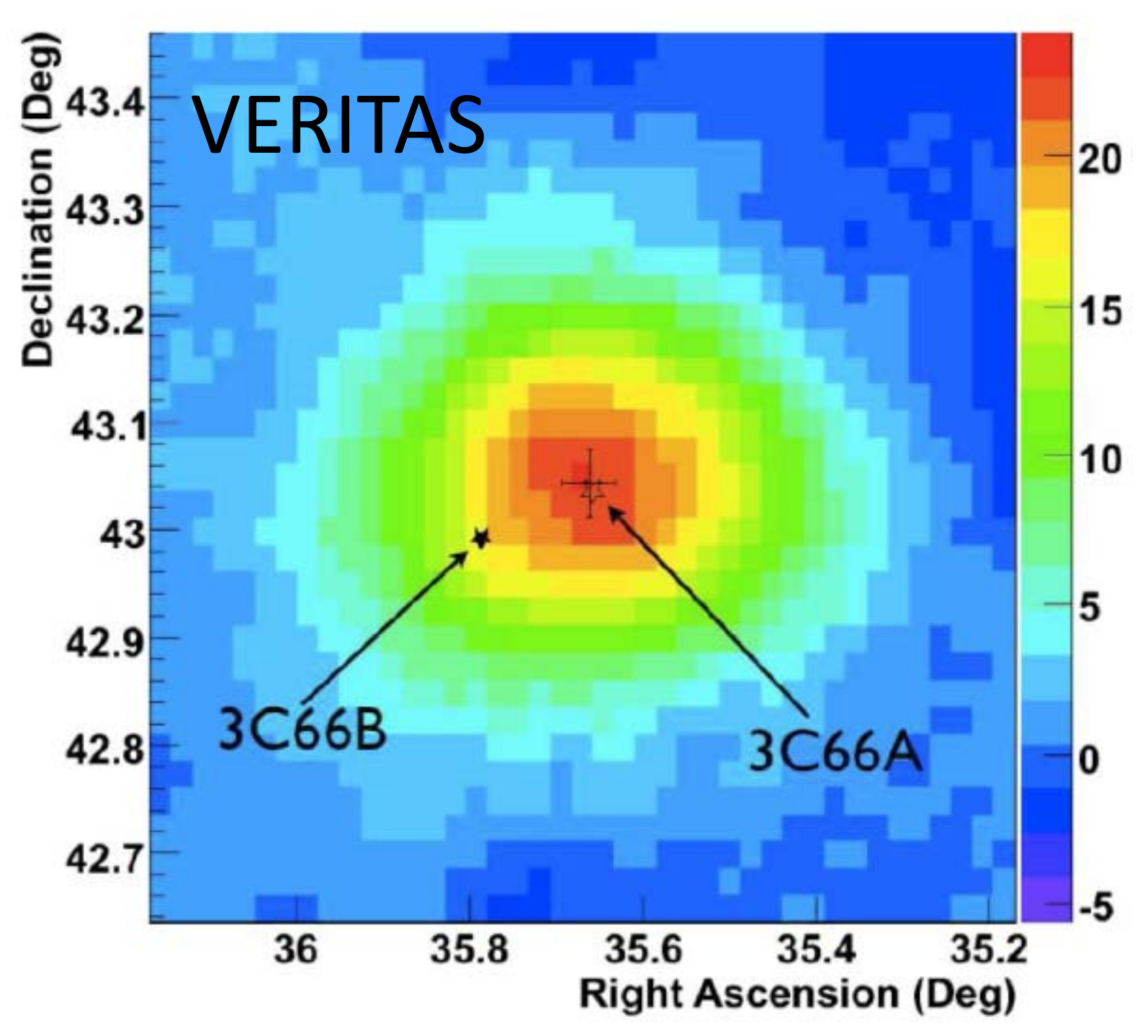}
  \caption{Smoothed significance map derived from
  VERITAS observations of 3C 66A. The cross
  shows the VERITAS localization of the VHE
  emission. The open star indicates the position
  of 3C 66A and the closed star indicates the
  position of 3C 66B.}
\end{figure}

The third IBL, and the latest blazar, to be discovered
at very high energies 
by VERITAS is PKS 1424+240.
This source, with an unknown redshift,
was first detected in gamma rays by Fermi
\cite{Cygnus_Fermi}.
The VERITAS detection came from 14 h of
data taken in Spring 2009 \cite{PKS1424_VERITAS}.
The source was relatively weak at VHE energies,
at a flux level $\sim$2\% Crab Nebula above
200 GeV.
PKS 1424+240 is the first VHE discovery motivated
by Fermi observations. 
As shown in Figure~5, joint analysis of
the VERITAS and Fermi data provides constraints
on both the redshift of the source as well as the
inverse-Compton model parameters
\cite{PKS1424_VERITAS_Fermi}.

\begin{figure}[!t]
  \centering
  \includegraphics[width=3.2in]{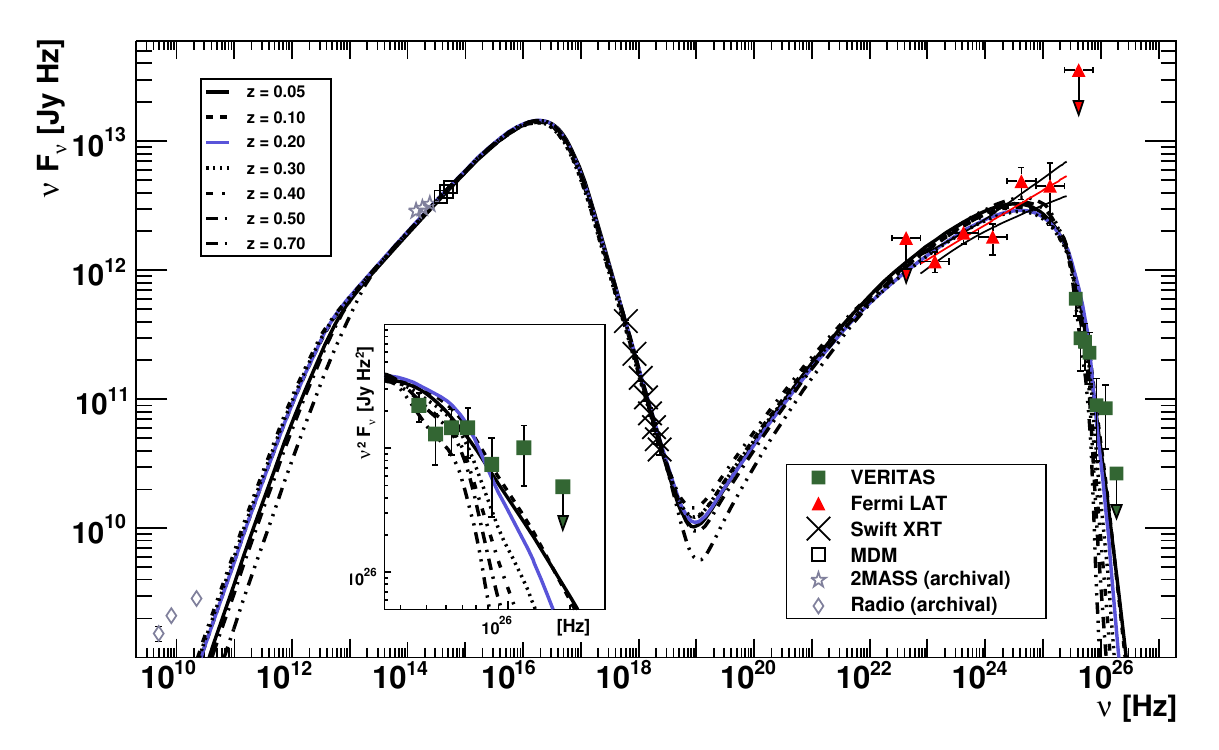}
  \caption{
   Spectral energy distribution of PKS 1424+240,
   from \cite{PKS1424_VERITAS_Fermi}.
   The lines show SSC=model fits assuming different
   redshifts. 
   The inset shows a zoom of the VERITAS data.
   The Fermi data are presented with their corresponding
   power-law fit and one standard deviation uncertainty.}
\end{figure}

Additional blazar papers presented
at this conference
by VERITAS describe variability of the
VHE emission from 1ES 1218+304\cite{Imran} and
multiwavelength studies of Mrk 421, Mrk 501 and
1ES 2344+514 \cite{Grube}.
Results from the Whipple 10m blazar monitoring
program of five key sources were also presented
\cite{Pichel}.

\subsection{Radio Galaxies and Gamma-Ray Bursts}

Almost all of the AGN detected in the VHE gamma-ray
band are blazars, but there are also two
radio galaxies seen: M 87 and Centaurus A.
The fact that these objects are much closer to us 
than the blazars allows better resolution of their
structure.
M 87 is a giant radio galaxy in the Virgo cluster,
Misalignment of its jet relative to the 
line-of-sight to Earth permits imaging of 
the jet in the radio, optical and X-ray bands.
After its first detection by HEGRA \cite{M87_HEGRA},
M 87 has now been extensively studied
by VHE gamma-ray telescopes.

M 87 was first detected by VERITAS in 2007, at
flux level of $\sim$ 2\% Crab Nebula
\cite{M87_VERITAS}.
In this epoch the source exhibited relatively
little variability.
In February 2008, however, strong flaring in gamma rays
was detected during a joint observation
campaign involving the VLBA and
the VHE instruments
VERITAS, MAGIC, and HESS \cite{Wagner,M87_joint}.
During this flaring, Chandra revealed
the nucleus of M 87 to be active
in the X-ray band, providing evidence
that the TeV photons are emitted from the
core of M 87.
In 2009, M 87 is apparently in a relatively
low state; $\sim$20 h of observation
by VERITAS
yielded only a marginal detection \cite{Hui}.
Further multiwavelength efforts are likely
needed to provide clear insight into the
acceleration and emission mechanisms of
this fascinating source.

Gamma-ray bursts (GRBs) are the most powerful cosmic explosions
known,
with complex acceleration mechanisms that likely
involve shocks in a highly relativistic jet.
To date, no convincing evidence of VHE emission
from GRBs has been presented, although GeV
photons have been detected by both EGRET and now
Fermi-LAT.
The targeting of GRBs is very high priority for
VERITAS. 
Since 2006, 31 GRBs have been observed.
The response time of VERITAS to GRB alerts is
excellent, with typical delays of two to four
minutes from the beginning of the burst and
92 s as the best case \cite{Galante2}
So far, no detections have been made, but the
future looks promising for an upgraded VERITAS
with improved sensitivity and lower energy 
threshold \cite{Otte}.

VERITAS also reported results from observations
of the radio galaxies NCG 1275 and 3C 111
and the Coma cluster of galaxies \cite{Galante1}.

\section{Galactic Sources}

The Galaxy is a rich source of high-energy
gamma-ray emission, with 90\% of the astrophysical
photons seen at GeV energies corresponding to diffuse
emission in the Galactic plane.
To date, we have four types of Galactic
objects at TeV energies: 
pulsar wind nebulae (PWN),
supernova remnants (SNRs), binary systems,
and unidentified sources.
In these objects we study the acceleration
of electrons and protons in shock fronts, colliding
winds, superbubbles, etc., with a primary goal
of pinning down the origin of cosmic rays.

The observation of Galactic sources is a
high priority for VERITAS.
In addition to the Sky Survey discussed earlier, there is
a second key science project focused on PWN and
SNRs \cite{Humensky}.
Here we report on new detections by VERITAS of
four Galactic sources.

\subsection{G54.1+0.3 and G106.3+2.7 (Boomerang)}

The supernova remnant G54.1 +0.3 has many similarities
to the Crab Nebula, with an X-ray jet and torus
being observed around the pulsar PSR J1930+1852.
With an age of $\sim$2,900 years and
a spin-down luminosity of $\dot{\rm E} \sim 1.2 \times 10^{37}$\,erg/s,
this remnant/PWN is a likely candidate for VHE gamma-ray
emission.
The presence of a nearby molecular cloud as a possible
target material for VHE cosmic rays provides further observational
motivation.

Following a hint of a signal from moonlight data taken in
2007, G54.1+0.3 was observed by VERITAS for 31\,h in 2008,
yielding a solid detection at the 7.0$\sigma$ level.
The VHE emission is consistent with a point source
at the pulsar location.
The preliminary flux level is $\sim$3\% Crab Nebula
above 1 TeV.
As shown in Figure~6,
the preliminary differential spectral index is
$\Gamma = 2.3 \pm 0.3_{\rm stat}
\pm 0.3_{\rm sys}$.

\begin{figure}[!t]
  \centering
  \includegraphics[width=3.0in]{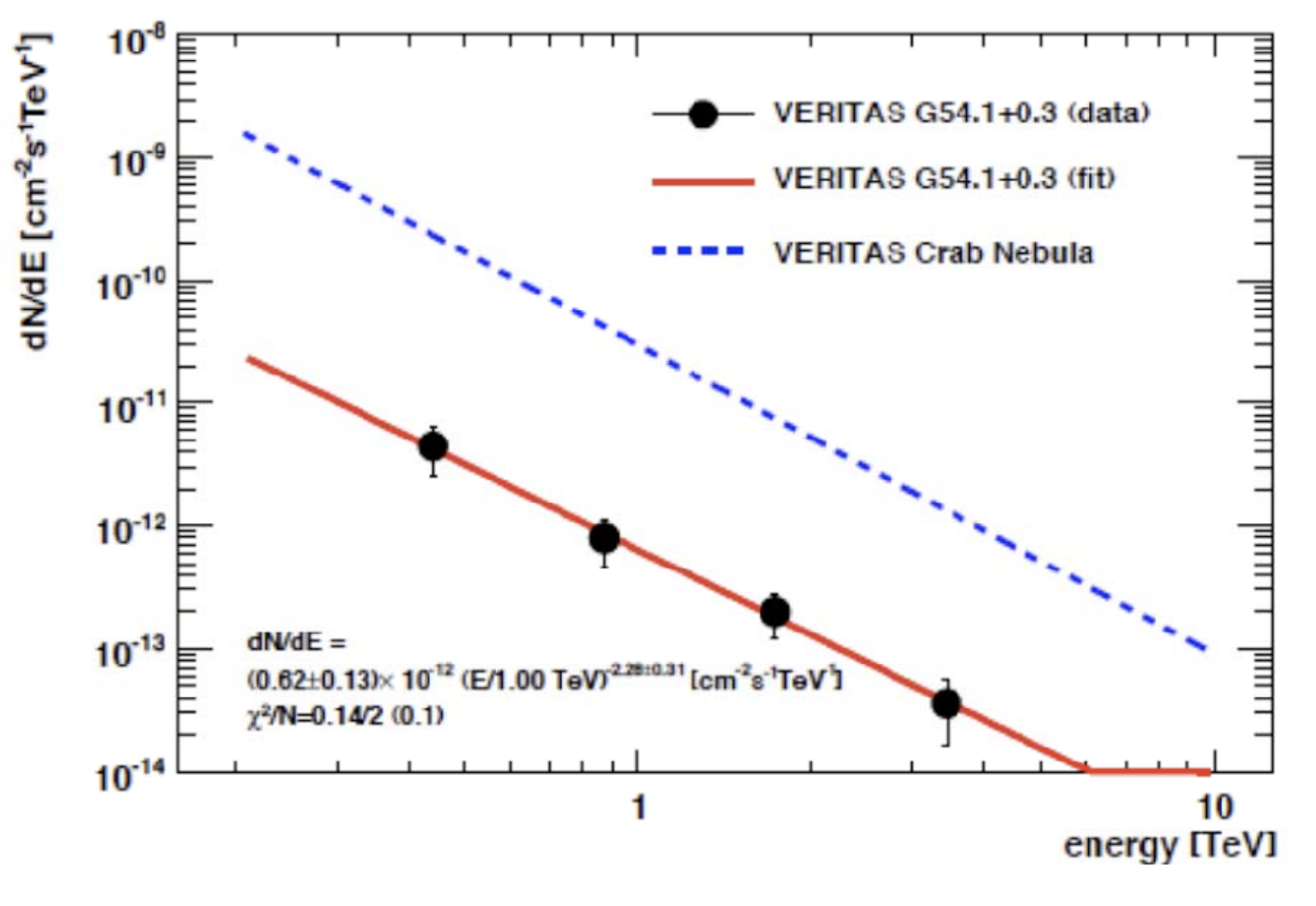}
  \caption{Preliminary differential energy spectrum
  as measured by VERITAS for the VHE gamma-ray
  emission from SNR G54.1+0.3.  The spectrum
  is well-fit by a power-law form with
  spectral index
  $\Gamma = 2.3 \pm 0.3 ({\rm stat})
  \pm 0.3 ({\rm sys})$.}
\end{figure}

The supernova remnant G106.3+2.7 is 
part of a complex system that may have been
created by a supernova explosion occurring 
in a previously existing HI bubble
\cite{Kothes}.
The energetic pulsar associated with this
system, PSR J2229+6114,
has an estimated age of $\sim$10,000 years and
a spin-down luminosity of 
$\dot{\rm E} \sim 2.2 \times 10^{37}$\,erg/s.
The SNR is within the error box of the
EGRET source 3EG J2227+6112, and the pulsar
appears on the Fermi Bright Source List
\cite{Cygnus_Fermi}.
Milagro reported $>$10 TeV emission from the general
region \cite{BSL_Milagro} with a large error
box $\sim 1^\circ$ in diameter.

The VERITAS detection of VHE emission came
from 33 h of observations carried out in 2008 that
resulted in a post-trials significance of
6.0$\sigma$ and an integral gamma-ray flux
level of $\sim$5\% Crab Nebula above 
1 TeV \cite{Boomerang_VERITAS}.
As shown in Figure~7, the
VHE emission is clearly extended, spanning
a region approximately $0.4^\circ$ by $0.6^\circ$
in size.
However, the peak of the emission is clearly
displaced from the pulsar and instead overlaps with
a region of high CO density.
The measured VHE spectrum,
with differential spectral index
$\Gamma = 2.3 \pm 0.3_{\rm stat}
\pm 0.3_{\rm sys}$, is relatively hard and
is consistent with a power-law form up
the Milagro energy of 35 TeV.
The spectrum and the observed morphology of the
source support a possible hadronic origin
for the VHE emission.

\begin{figure}[!t]
  \centering
  \includegraphics[width=3.4in]{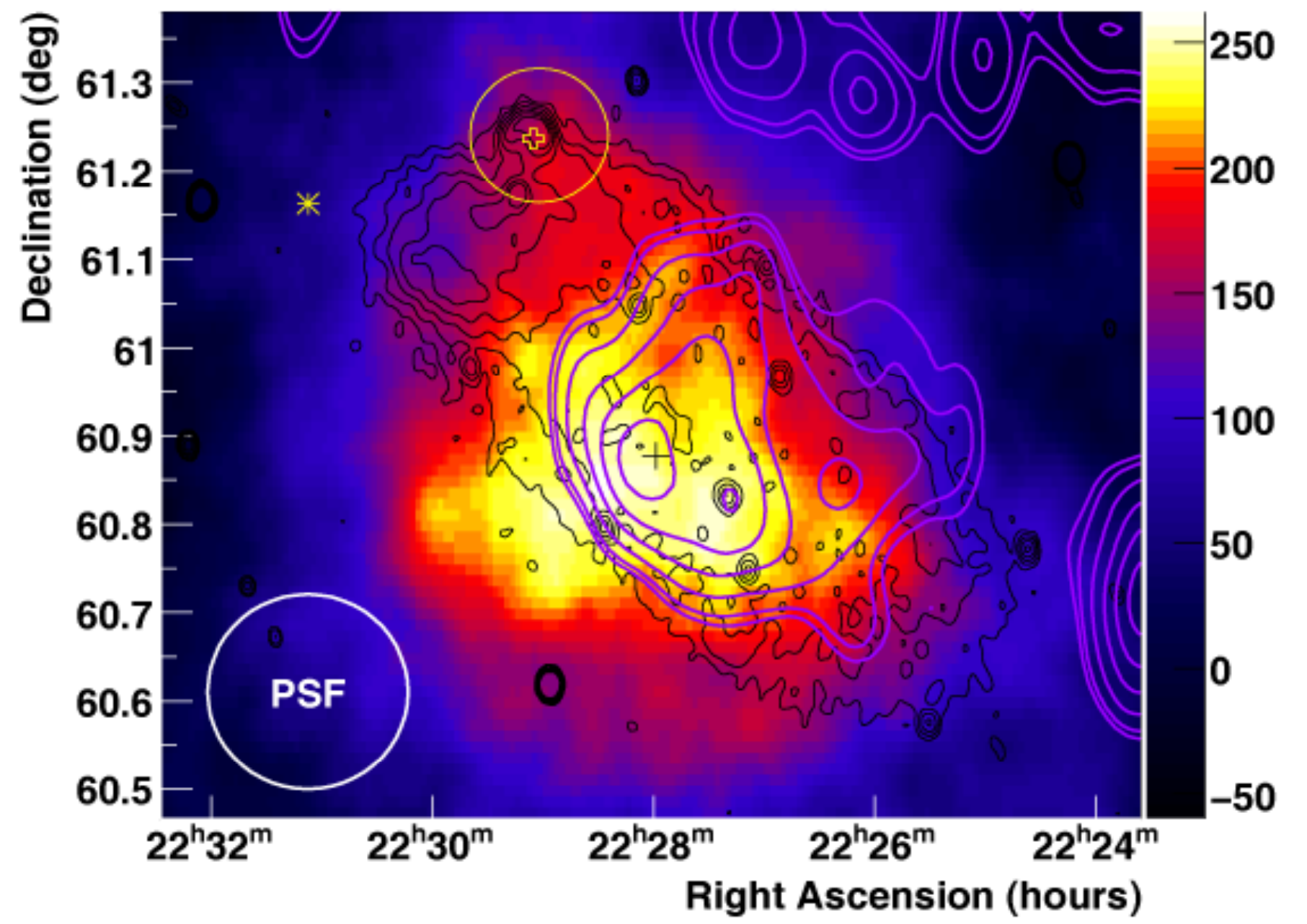}
  \caption{Sky map of VHE gamma-ray emission
   from G106.3+2.7 as measured by VERITAS.
   The color scale indicates the excess gamma-ray
   events from the region; the centroid of the
   VHE emsision is indicated by the black cross.
   The yellow circle indicates the 95\% error
   contour for the Fermi source
   OFGL J2229.0+6114 and the white circle represents
   the gamma-ray point spread function (68\%
   containment) of VERITAS.  Radio contours
   and CO emission contours are indicated by the
   black and magenta lines, respectively.}
\end{figure}

\subsection{Other Supernova Remnants}

The SNRs Cassiopeia A (Cas-A) and IC 443 are now
well established VHE gamma-ray sources.
Observations of Cas-A by VERITAS in 2007 totalled
22 h and yielded a clear detection at the
8.3$\sigma$ statistical level \cite{Humensky2}.
The integral gamma-ray flux is
$\sim$3.5\% Crab Nebula above 1 TeV.
The VERITAS energy spectrum is well fit by a power-law
form with differential spectral index
$\Gamma = 2.6 \pm 0.3_{\rm stat}
\pm 0.2_{\rm sys}$ and there is no
indication of a cut-off at high energy.
There is also no evidence for any source
extension \cite{CAS_A_VERITAS}.

The emission of VHE gamma rays from IC 443
was first reported by MAGIC and VERITAS
in April 2007 at the VERITAS First Light Celebration.
In their subsequent paper, MAGIC reported
a 5.7$\sigma$ detection of the source, corresponding
to an integral flux of $\sim$2.8\% Crab Nebula
above 300 GeV \cite{IC443_MAGIC}.
The initial VERITAS data set on IC 443,  
of $\sim$16 h taken in 2007 with three telescopes ,
was augmented in 2008 by $\sim$21 h taken 
with four telescopes.
The overall data set yields a post-trials
statistical significance of 7.5$\sigma$ and
an integral flux of $\sim$3.2\% Crab Nebula,
consistent with the MAGIC result.
However, the VERITAS data show for the first
time that the gamma-ray emission is extended,
with a characteristic fitted
two-dimensional Gaussian radius of
$0.16^\circ$ \cite{IC443_VERITAS}.
The VHE emission also overlaps with a dense
CO molecular cloud.
Further observations at both GeV and TeV
energies are needed to unambiguously determine
whether the primary particles accelerated in
IC 443 are protons or electrons (or some
combination).

\subsection{Unidentifieds and Other Galactic Sources}

The W44 SNR has long been a candidate source of
TeV gamma-ray emission.
HESS detected two unidentified sources,
HESS J1857+026 and HESS J1858+020,
in the region, but not W44 itself \cite{W44_HESS}.
Based on 13 h of observations taken in
2008, VERITAS confirms the HESS results.
HESS J1857+026 is detected at the
5.6$\sigma$ significance level,
with the VERITAS data showing a comparable
spectrum and position to HESS.
HESS J1858+20 is marginally detected
at the 3.4$\sigma$ level.
W44 is not detected in the VERITAS data
and an integral flux upper limit of
$< 2.0$\% Crab Nebula is obtained.

HESS J0632+057 was discovered 
during HESS observations of the Monoceros Loop
in 2004 and 2005 \cite{HESS_J0632}.
At that time, the source was at a flux
level of $\sim$3\% of the Crab Nebula.
As one of only two unidentified sources in the
Galactic plane consistent with
being point-like, 
HESS J0632+057 was logically postulated
to be a binary system \cite{Hinton}.
VERITAS observed the source over a three
year period for a total
of $\sim$30 h, with no strong
evidence for VHE gamma-ray emission
\cite{Maier2,J0632_VERITAS}.
From these data, an integral flux upper limit from
the source of
$< 1.1$\% Crab Nebula above 1 TeV 
is obtained.
The VERITAS observations exclude the possibility
of steady VHE emission from HESS J0632+057
at $> 99.99$\% C.L..
This strongly supports the source being
variable, and it could well be a binary.

Other VERITAS Galactic source 
results reported at this conference 
include long-term studies of the TeV
binary LS I +61 303 \cite{Holder2},
observations of
globular clusters \cite{McCutcheon},
Geminga \cite{Finnegan},
X-ray binaries \cite{Guenette1},
magnetars \cite{Guenette2},
and a forbidden velocity wing \cite{Holder3},
and a search for bursts of gamma rays
from the Crab Pulsar \cite{Schroedter2}.

\section{Dark Matter Searches}

The nature of dark matter is one of the most compelling
mysteries in physics and astronomy today.
Particle dark matter candidates, {\em e.g.} the
weakly interacting massive particle (WIMP),
predict unique high energy gamma-ray signatures.
For example, the WIMP self-annihilation signal
is expected to yield a continuum 
gamma-ray spectrum that terminates at the WIMP
mass.
Depending on the particle physics, there may also
be gamma-ray line emission.

The search for dark matter is a key science
project of VERITAS, encompassing
$\sim$6\% of the observing time.
To carry out a comprehensive search of
astrophysical systems that are likely to
contain a preponderance of dark matter,
we targeted a variety of objects, including
nearby dwarf galaxies ({\em e.g.} Draco, Ursa Minor),
local galaxies ({\em e.g.} M32, M33),
globular clusters ({\em e.g.} M5), and
galaxy clusters ({\em e.g.} Coma).
So far, 
no clear gamma-ray
signals are detected from any of 
the dark-matter
candidate sources and strong limits are placed
on the gamma-ray emission from 
seven targets \cite{Wagner2}.

\section{IceCube Hotspot}

The large IceCube neutrino telescope is currently
under construction at the South Pole, and
the partially-completed detector is already
carrying out searches for astrophysical
sources of VHE neutrinos.
The all-sky neutrino map from the 22-string
IceCube detector revealed a hotspot
(excess of 7.7 events) at the sky position
($\alpha$, $\delta$) = (10h13m30s, +11d22m30s)
\cite{Montaruli}.
Information about this hotspot was conveyed
to VERITAS and Director's Discretionary Time
was used to carry out observations of it.
VERITAS observed the IceCube hotspot for
2.5 h in April 2009 in moonlight conditions. 
No signal was detected
and an upper limit on the integral flux of 
gamma rays of $< 4.0$\% Crab Nebula
above 1 TeV was obtained.

\section{Future}

Given the excitement of the field of gamma-ray astronomy,
it is natural to consider ways to improve the 
performance of VERITAS.
This goal is especially true in light of the
unique capabilities of Fermi, currently planned
to operate through 2013, at least, and very
likely beyond that.
Given this timetable, it makes sense to consider
upgrade options that can be carried out on
a time scale of two to three years.

The VERITAS upgrade program has three stages.
The first stage, already accomplished, involved
improving the optical point spread function
through better mirror alignment
\cite{McCann} and the relocation of Telescope 1.
As shown in Figure~1, the first VERITAS telescope
was moved in Summer 2009 to increase the baseline
distances between it and the other telescopes.
As discussed in Section~1,
this change in the layout of VERITAS has had
a significant impact on its performance. 
A point source with a flux of 1\% Crab Nebula can
now be detected in under 30 h.

The second stage of the upgrade program is
aimed at further improving the sensitivity and
extending the reach of VERITAS to lower energies
\cite{Otte}. We are proposing to upgrade each
VERITAS camera by replacing the existing
PMTs with ones having 
higher quantum efficiency.
A new topological telescope trigger system
is also envisioned \cite{Schroedter}.
Possibilities for a future third upgrade stage
include an automatic mirror alignments system and
an additional telescope (T5).

\section{Summary}

The four-telescope VERITAS array 
is operating extremely well ($> 95$\% uptime)
and with excellent sensitivity.
Based on two years of quality data, VERITAS
presented many new results at this meeting,
including:
\begin{itemize}
\item the discovery of gamma-ray emission from the
starburst galaxy M 82,
\item stringent limits on source fluxes from the
Galactic plane sky survey,
\item the detection of five new blazars
(1ES 0806+524, W Comae, 3C 66A, RGB 0710+541,
and PKS 1424+240),
\item correlated multiwavelength emission from
the radio galaxy M 87 (with MAGIC and HESS),
\item the detection of two new SNRs/PWN 
(G106.3+2.7 and G54.1+0.3),
\item detailed studies of the supernova remnants
IC 443 and Cas-A, and
\item limits on the annihilation of dark matter to
VHE gamma rays in seven astrophysical targets.
\end{itemize}

The current VERITAS sky map is shown in
Figure~8. 
VERITAS has detected 21 VHE gamma-ray
sources, eight previously not seen by
other instruments.
An upcoming upgrade program will further improve
the performance of VERITAS, ensuring that it
remain a premier gamma-ray observatory well
into the next decade.

\begin{figure}[!t]
  \centering
  \includegraphics[width=3.25in]{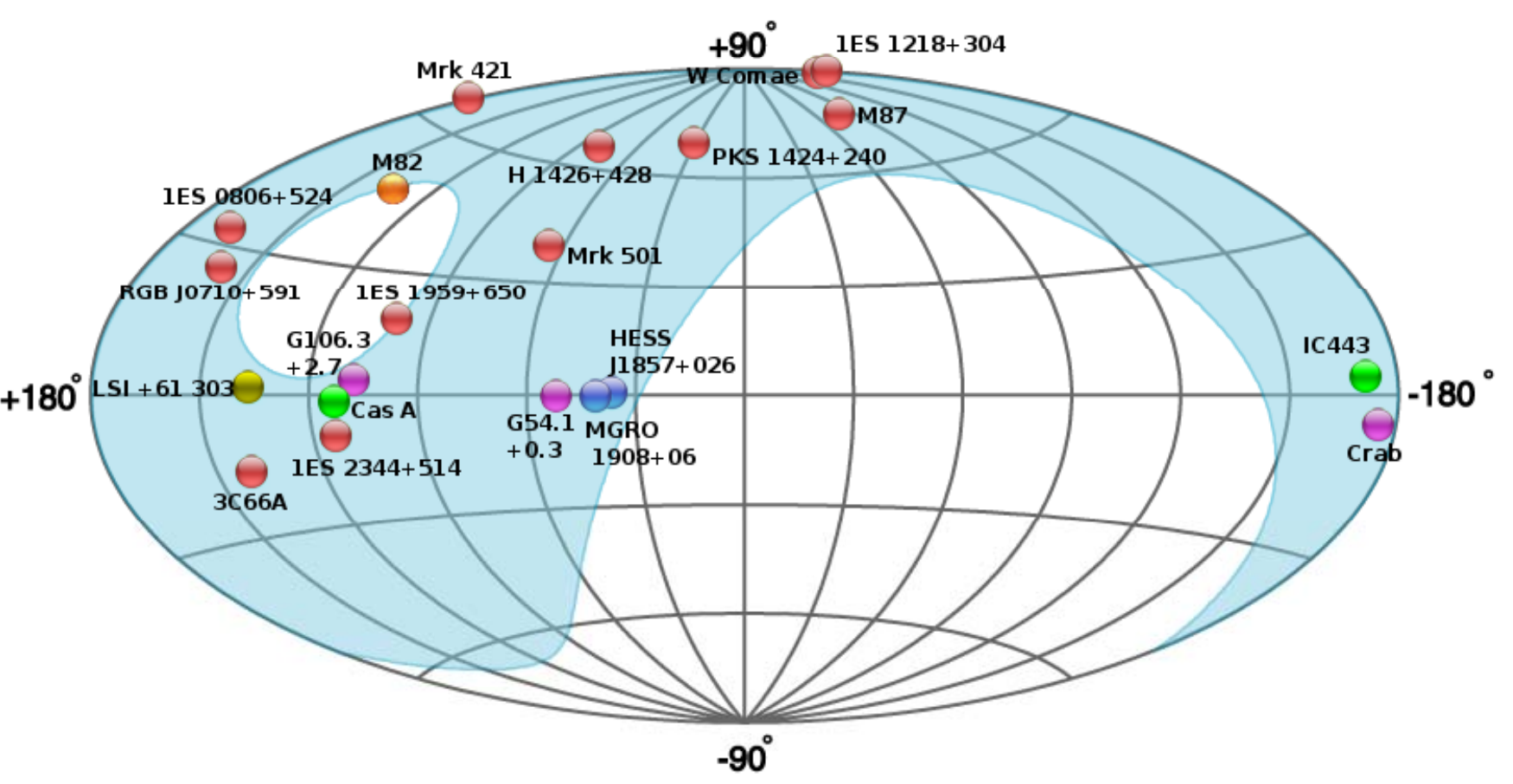}
  \caption{Skymap of VERITAS VHE gamma-ray
   source detections, as of July 2009. The
  various source classes are:
  AGN (red), binary systems (yellow),  
  pulsar wind nebulae (purple),
  starburst galaxies (orange), 
  supernova remnants (green), and
  unidentified sources (blue).}
\end{figure}

\section{Acknowledgments}

This research is supported by grants from  
the U.S. National Science Foundation, 
the U.S. Department of Energy,
and the Smithsonian Institution, by NSERC in
Canada, by Science Foundation Ireland, and by STFC in the U.K.  We acknowledge
the excellent work of the technical support staff at the FLWO and the
collaborating institutions in the construction and operation of the instrument.

\end{document}